# Frequency pulling and mixing of relaxation oscillations in superconducting nanowires

Emily Toomey, Qing-Yuan Zhao, Adam N. McCaughan, Karl K. Berggren*

*Massachusetts Institute of Technology, Department of Electrical Engineering and Computer Science, Cambridge, MA, 02139*

*correspondence to: berggren@mit.edu

**ABSTRACT**

Many superconducting technologies such as rapid single flux quantum computing (RSFQ) and superconducting quantum interference devices (SQUIDs) rely on the modulation of nonlinear dynamics in Josephson junctions for functionality. More recently, however, superconducting devices have been developed based on the switching and thermal heating of nanowires for use in fields such as single photon detection and digital logic. In this paper, we use resistive shunting to control the nonlinear heating of a superconducting nanowire and compare the resulting dynamics to those observed in Josephson junctions. We show that interaction of the hotspot growth with the external shunt produces high frequency relaxation oscillations with similar behavior as observed in Josephson junctions due to their rapid time constants and ability to be modulated by a weak periodic signal. In particular, we use a microwave drive to pull and mix the oscillation frequency, resulting in phase locked features that resemble the AC Josephson effect. New nanowire devices based on these conclusions have promising applications in fields such as parametric amplification and frequency multiplexing.

# I. INTRODUCTION

Relaxation oscillators have been used to model a wide variety of nonlinear behavior found in biological and physical systems; for instance, cardiac rhythms [1]and modulated semiconductor lasers [2] have both been described by relaxation oscillation dynamics. While in the most basic sense, relaxation oscillations are comprised of a nonlinear element and a feedback cycle, it was observed early on that another property unique to relaxation systems is that their oscillation frequencies may be altered by the application of a weak periodic drive[3],[1]. As a result, both the relaxation process and its response to external stimuli are vital to the characterization of a complete system.

Superconductors represent an ideal platform for studying and manipulating these types of oscillatory phenomena. In addition to having rapid nonlinear switching dynamics, their response changes with temperature, current density, and the application of magnetic fields. This tunability allows for the effects of external drives to be observed. Perhaps one of the strongest manifestations of these oscillations in superconductors is the AC Josephson effect, in which locking between a periodic drive and the junction's sinusoidal current-phase relationship produces zero-slope regions known as Shapiro steps in the current-voltage curve at intervals of $V_n = nhf/2e$, where $f$ is the driving frequency, $h$ is Planck's constant, and $e$ is the electronic charge [4],[5]. This relationship between frequency and voltage has been exploited in technologies such as superconducting analog-to-digital converters (ADCs) in which a Josephson junction produces single flux quantum (SFQ) pulses at a frequency corresponding to the applied voltage[6]. Another form of superconducting weak-link known as the Dayem bridge proved to have similar AC behavior, with additional steps appearing at subharmonic values of the Shapiro voltage due to its multivalued, periodic current-phase relationship[7],[8],[9]. Work by Calander et al. demonstrated that subharmonics can also occur in inductively shunted tunnel junctions by injection locking of relaxation oscillations; as the strength of the locking signal increased, the oscillation frequency was pulled towards the drive frequency, and mixing products were observed[10].

Unlike the tunneling of Josephson junctions, superconducting nanowires are governed by thermal transitions into the normal state as a result of Joule heating. Despite the lack of a sinusoidal current-phase relationship, they can support relaxation oscillations due to the nonlinear interactions between the resistive hotspot and the impedance of the external readout circuit, as previously demonstrated in superconducting nanowire single photon detectors (SNSPDs) [11],[12]. In work by Hadfield et al., oscillations in high inductance nanowires were attributed to the relaxation of a resistive hotspot through an inductor and resistive shunt. The appearance of Shapiro-like behavior upon microwave radiation was suspected to be the result of phase locking of these relaxation oscillations to the external drive [13].

Building on this suspicion, we present a study of the nonlinear dynamics in a resistively shunted nanowire, as summarized in Figure 1. The experimental RF voltage output reveals distinct oscillations which are in agreement with electrothermal simulations. A simple analytical model based on hotspot relaxation is used to explain this behavior. By using microwave radiation, we demonstrate that the hotspot oscillations can mix with a weak external signal and be pulled towards its frequency until becoming locked.

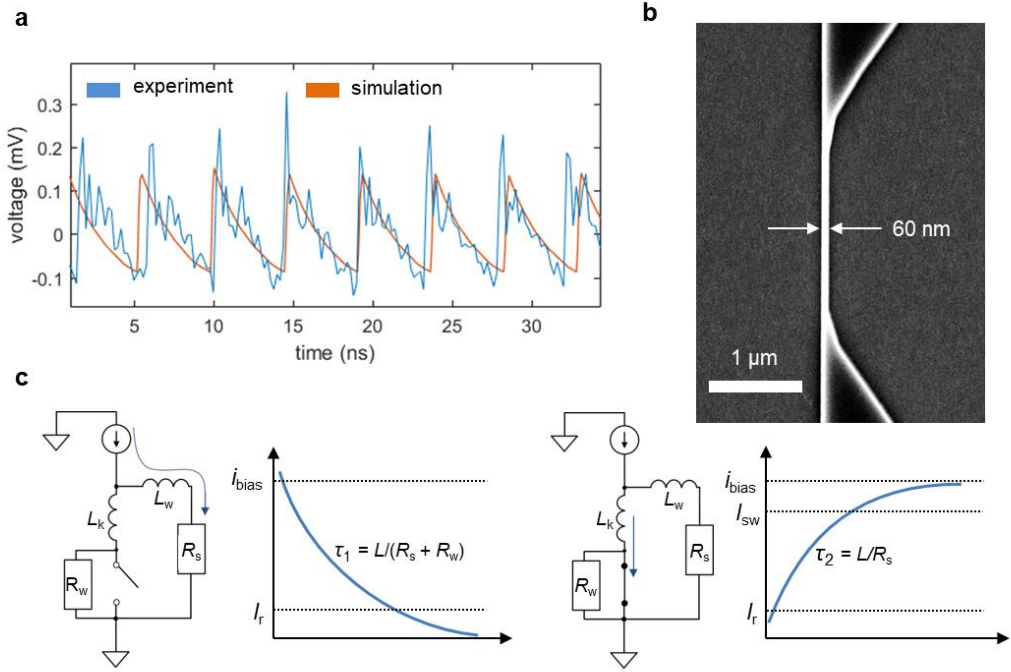

**FIG.1.** (a) RF voltage output of the shunted nanowire ($R_s$ = 10 Ω) when a bias current ramp is applied. The blue trace corresponds to experimental data, and the red trace is the output of electrothermal simulations. (b) Scanning electron micrograph of the tapered nanowire. (c) Illustration of a basic relaxation model. The bias current is first diverted to the shunt after the nanowire switches to the normal state, and then returns to the nanowire once superconductivity has been restored. The total inductance $L$ is the sum of the nanowire's kinetic inductance $L_k$ and the inductance of the wire bond connections $L_w$. In our experimental setup, $L_w$ dominates the total inductance.

## II. METHODS

The nanowires studied in this paper were fabricated from thick (∼40 nm) film niobium nitride deposited on Si substrates. Contact pads were defined by photolithography on evaporated Au and Ti. Superconducting nanowires were designed with tapers from 1 μm to a minimum width of 60 nm to minimize current crowding [14], and were written using electron beam lithography (125 kV Elionix) of hydrogen silsesquioxane resist (HSQ), followed by development in 25% TMAH and reactive ion etching in CF$_4$. The total length of the device (see Fig. 1b) was 30 μm, with a 60 nm wide center region about 1.5 μm long.

In order to shunt the nanowire, an external resistor of 10 Ω was placed in parallel less than 5 mm away on the printed circuit board (PCB). Aluminum wire bonds were used to make electrical connections to the contact pads and the shunt. All measurements were conducted in liquid helium at a temperature of 4.2 K, well below the critical temperature of the superconducting film (T$_c$ ~ 10.5 K). Electrical transport characterization for current-voltage measurements was achieved using a four-point scheme and a sinusoidal current bias at a sweeping frequency of 10-20 Hz. DC output voltages were sent through a low-noise preamplifier (SRS560) before being read out by an oscilloscope. Microwave modulation was achieved by applying 100-900 MHz sinusoidal signals to an external wire loop placed less than 5 mm over the sample and soldered onto an input port of the PCB. Relaxation oscillations were measured using a two-point configuration and amplifying the RF voltage output of a bias-tee.

To explain the observed dynamics, we considered a basic relaxation oscillation model from the perspective of a shunted nanowire. Figure 1c depicts its two time domains. After the nanowire first switches into the normal state, the bias current $i_{bias}$ is diverted to the shunt with a time constant of $\tau_1 = L/(R_s + R_w)$, where $L$ is the inductance, $R_s$ is the shunt resistance, and $R_w$ is the resistance of the nanowire, represented by the hotspot resistance. Once the current through the nanowire falls below the retrapping current $I_r$ and allows the nanowire to return to the superconducting state, the bias current is diverted back from the shunt with a time constant $\tau_2 = L/R_s$ until the switching current $I_{sw}$ is reached and the wire becomes normal again. Thus, the total period of a single relaxation oscillation may be approximated as:

$$T = -\tau_1 \ln\left(\frac{I_r}{I_{sw}}\right) - \tau_2 \ln\left(\frac{i_{bias} - I_{sw}}{i_{bias} - I_r}\right) \qquad \text{Eq. 1}$$

Since $R_s$ tends to be much less than $R_w$, the duration set by $\tau_2$ $\tau_2$ is expected to dominate the oscillation frequency, as was found in similar modeling of relaxation oscillations in hysteretic Josephson bridge contacts[15].

## III. RESULTS AND DISCUSSION

The expression given in Eq. 1 indicates that the frequency of the oscillation is a function of both the bias current and the inductance between the hotspot and the shunt resistor. Figure 2 shows our experimental data to investigate these dependences, and compares the results with the simplified expression given in Eq. 1 and with electrothermal simulations conducted in SPICE[16], [17]. The measurement was performed using a DC battery source in series with a 10 kΩ resistor to apply a steady current bias to the shunted device, and the oscillation frequency was extracted from the FFT of the RF voltage output of the bias-tee. The process was repeated for two different inductances in series with $R_s$, which was set by changing the distance between the shunt and the nanowire and thus altering the wire bond length. The hotspot resistance $R_w$ was used as a fitting parameter for both the electrothermal simulation and the relaxation oscillation model, and was ultimately set equal to 500 Ω (corresponding to a length of about 500 nm over the 60 nm wide region). For both models, the best-fit series inductance changed by roughly 25% between the two experiments with different wire bond lengths, revealing the significant impact of path inductance on the nonlinear response of the shunted system. In addition to highlighting the influence of series inductance on the overall dynamics of the shunted nanowire, the agreement between the experimental data and the two models indicates that the underlying mechanism of oscillation is indeed the electrothermal feedback between the hotspot and the shunt, rather than the nonlinearity of the superconducting current-phase relationship as in a Josephson junction.

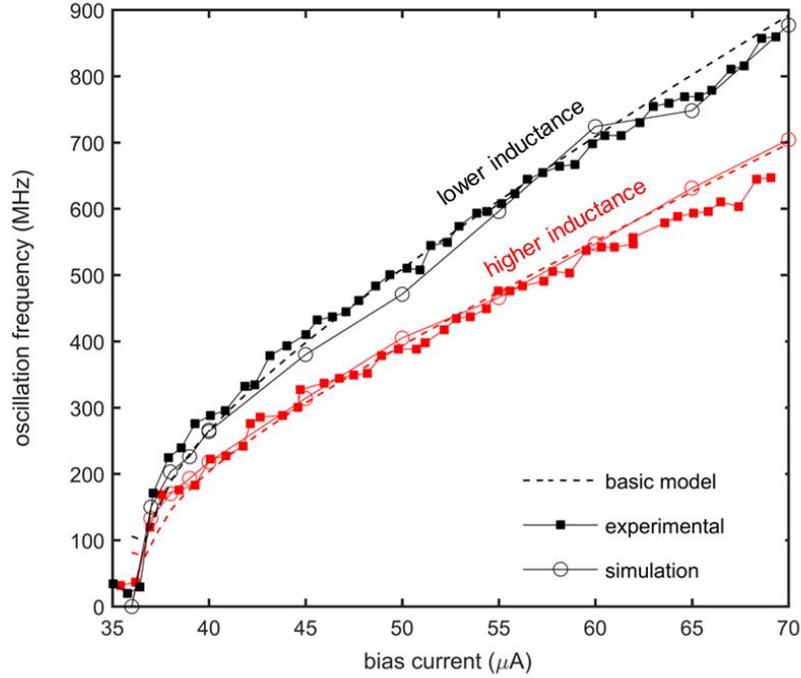

**FIG.2.** Relationship between $I_{\text{bias}}$ and the oscillation frequency for two different series inductances (red and black curves). The experimental data is compared to both electrothermal simulations and the basic relaxation oscillation model in Eq. 1. Parameters used to fit the red curve: (basic model) $L$ = 20 nH, (simulation) $L$ = 25 nH. Parameters used to fit the black curve: (basic model) $L$ = 15.3 nH, (simulation) $L$ = 18.75 nH. For all fits, $R_s$ = 10 Ω and $R_{\text{wire}}$ = 500 Ω.

By reaching a maximum oscillation frequency close to the GHz range, such rapid oscillations can affect the shape of the current-voltage characteristics (IVCs) conducted from a slow measurement, analogous to the key role played by fast Josephson oscillations in the IVC of a Josephson junction. In addition to impacting the frequency of these oscillations as described in Eq. 1, we observe that the shunt resistor significantly controls the amount of thermal dissipation in the nanowire as indicated by the degree of hysteresis. Figure 3a-b show the IVCs of the nanowire with and without the presence of an external shunt. Placing a resistance of 10 Ω in parallel with the wire resulted in complete suppression of hysteresis, suggesting a lack of sustained Joule heating [18], [19]. The non-hysteretic IVC was fit to the expression for the average DC voltage of an overdamped Josephson junction in which the contribution of capacitance can be neglected [20]:

$$V_{DC} = I_c R \sqrt{(I_{bias}/I_c)^2 - 1} \qquad \text{Eq. 2}$$

where $I_c$ is the critical current of the junction, $R$ is the total parallel resistance, and $I_{bias}$ is the current being supplied.

As illustrated in Fig. 3b, we observed good qualitative agreement between the experiment and the overdamped junction model when the nanowire is shunted with 10 Ω and when the model resistance is set equal to 7.8 Ω. The slight discrepancy in resistance is not surprising, since the total parallel resistance $R = (1/R_{wire} + 1/R_s)^{-1}$ is only expected to reach the full value of $R_s$ when the self-heating hotspot resistance has grown to the order of kΩ. Such a large hotspot impedance is avoided in a sufficiently shunted system if the resistance is low enough to divert the bias current after initial hotspot formation, limiting Joule heated expansion of the normal domain. Other simplifications in the model including a lack of noise and capacitance terms may also contribute to this deviation. Nonetheless, the qualitative agreement suggests a relationship between the shunted nanowire and an overdamped Josephson junction approximation, despite the presence of hotspots in the nanowire. Additionally, we observed that the shunt produced a narrowing in the width and an increase in the mean of the switching current distribution (Fig 3c). This observation has previously been explained in nanowires from the perspective of the Josephson junction tilted washboard potential model as an increased damping which reduces the likelihood of the phase particle entering the running voltage state by a single thermal fluctuation [21]. As a result, the mean switching current increases and the distribution becomes narrower due to the reduced impact of thermal fluctuations.

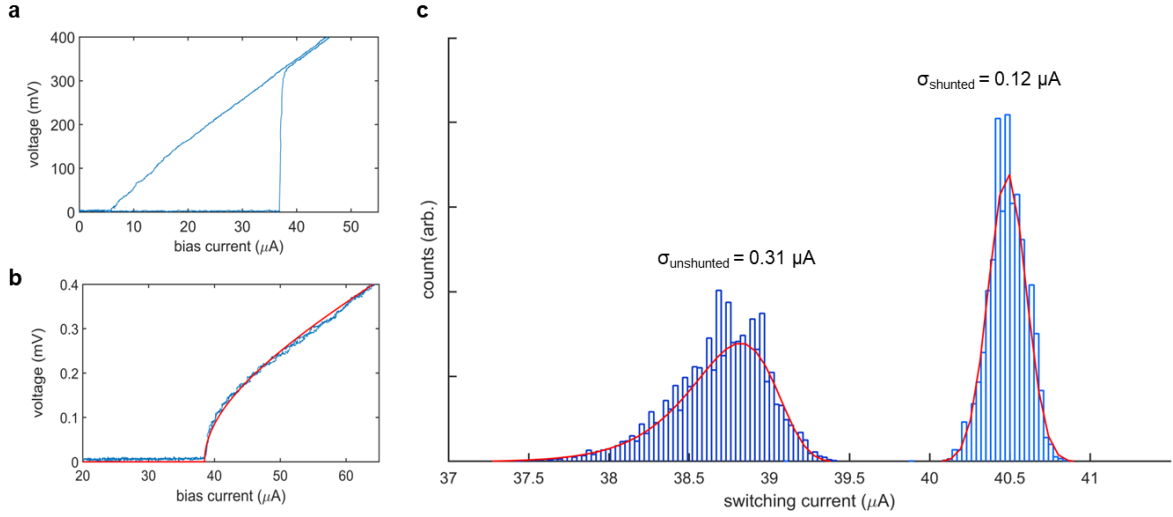

**FIG.3.** (a) Current-voltage characteristics of the unshunted nanowire. (b) Current-voltage of the nanowire when it is shunted with $R_s$ = 10 Ω. The red curve is a fit to an overdamped Josephson junction model (Eq. 1) with an $I_c$ = 38.5 μA and $R$ = 7.8 Ω. (c) Shunting the nanowire resulted in an increase in the mean and a narrowing in the width of the switching current distribution.

Although time-averaged IVCs mask the direct observation of relaxation oscillations, they may provide evidence of oscillation locking under the influence of external modulation, as in the AC Josephson effect. To conduct this investigation, four-point electrical characterization measurements were repeated while subjecting the shunted nanowire to external microwave radiation. As shown in Figure 4a, applying a 180 MHz sinusoidal drive to the coil antenna suppressed the nanowire's switching current and produced distinct steps in the IVC; the relative amplitudes of these steps appeared to have a Bessel-like relationship with the microwave power, eventually decaying to zero when the switching current was fully suppressed and the nanowire entered the normal state (shown in Fig. 4b).

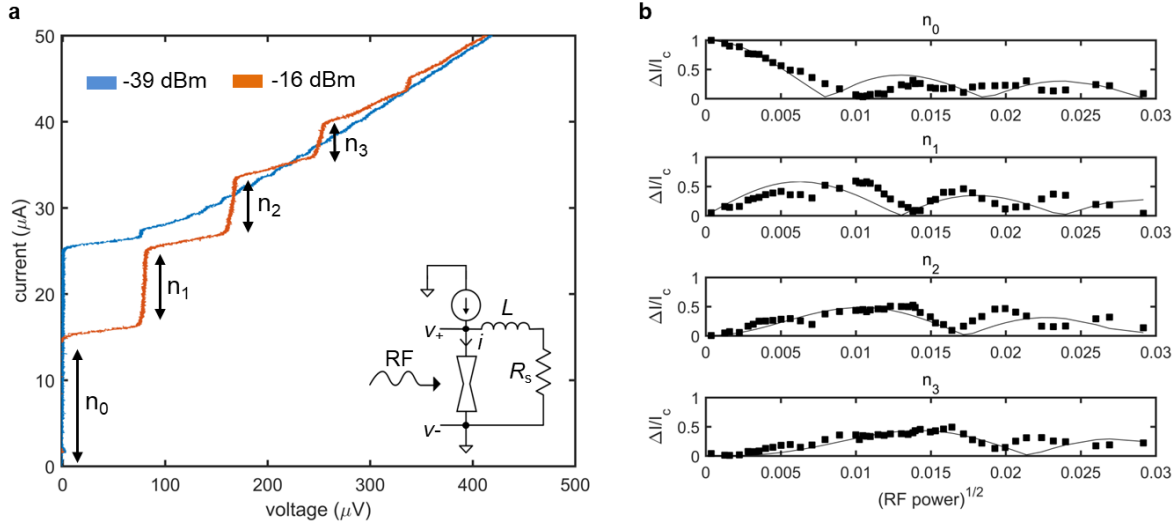

**FIG.4.** a) Steps appearing in the current-voltage characteristics with 180 MHz radiation at two different powers. b) Amplitudes of the first four current steps as a function of modulation power. Amplitudes are scaled relative to $I_c$, the magnitude of the $n_0$ step when no radiation is applied. The solid black curve shows fits to dynamical solutions for a Josephson junction, following the expression $\left| J_n\left(\frac{2e\alpha V}{hf}\right) \right|$ as described in [22], where $\alpha$ = 1e-4.

This phenomenon is similar to the Shapiro effect in Josephson junctions. Indeed, fits to the step amplitudes as a function of modulation power can be made using Josephson junction dynamical equations, as previously done for superconducting Dayem bridges[22]. In this case, step amplitudes follow the amplitudes of the $n^{th}$ order Bessel functions $J_n(2e\alpha V/hf)$, where $V$ is the magnitude of the external modulation and $\alpha$ is a fitting parameter describing the coupling loss between the coil and the nanowire in order to convert from the known output power to the unknown voltage delivered to the nanowire.

Despite this similarity, the steps occur at voltage intervals roughly 200 times the expected Shapiro voltage for a 180 MHz drive, a disparity that was also noted in the work by Hadfield et al. [13]. As a result, it is clear that this behavior does not stem from a Josephson-like current-phase relationship, but rather a different source of oscillation phase locking.

Frequency mixing and pulling of inductively shunted tunnel junctions to a weak periodic drive have previously been explored as a means of phase locking; the strong broadband oscillations of inductively shunted Josephson junctions were seen to mix with an injected narrowband signal, eventually producing a frequency spectrum peak at the injection frequency when the modulation power was sufficiently increased[10]. Figure 5 summarizes our search for similar dynamics in the resistively shunted nanowire. This investigation was conducted by examining the frequency spectrum of the RF voltage output at a steady bias current while subjecting the nanowire to increasing powers of microwave radiation. In addition to peaks appearing at the relaxation oscillation frequency ( $f_r$ = 504 MHz) and the drive frequency ( $f_d$ = 320 MHz), peaks were also observed at mixing products $f_r \pm m(f_d - nf_r)$. For instance, a peak at 690 MHz represents the mixing product of n = 1 and m = 1, or $2f_r - f_d$. Furthermore, as the magnitude of the driving signal increased, the relaxation oscillation peak was pulled towards the driving frequency, and the mixing products shifted in relation to the new $f_r$. This phenomena was also observed at a bias condition where $f_r$ was less than $f_d$ (194 MHz vs. 330 MHz), demonstrating that the oscillation frequency can be pulled in either direction.

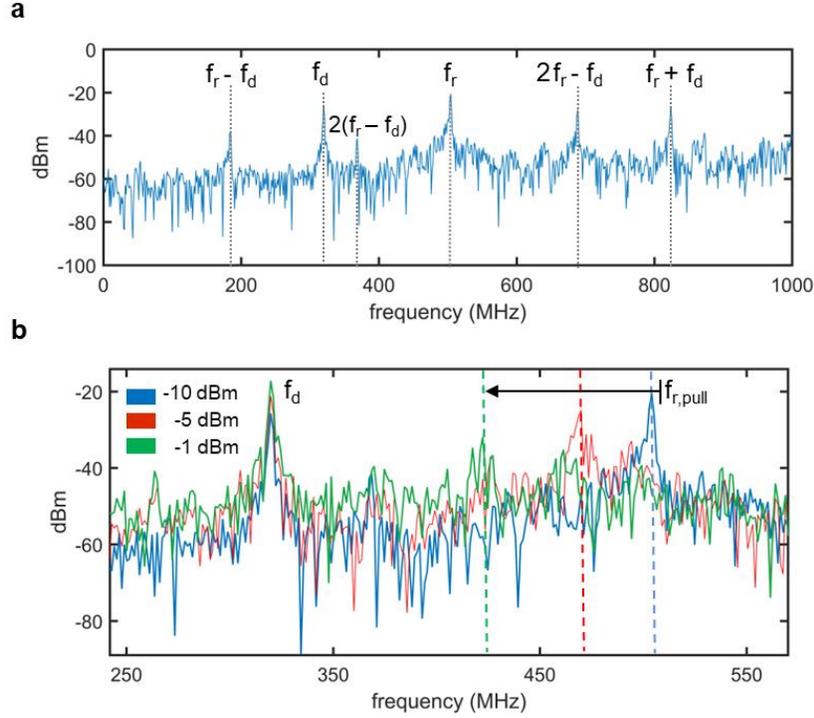

**FIG.5.** a) FFT of voltage output with an applied radiation of 320 MHz and a constant bias current of 60 μA. Peaks indicate the resonance of the relaxation oscillations and the driving frequency, as well as their mixing products. (b) Enlarged view of frequency spectrum while varying applied microwave power. As the power increases, the relaxation oscillation frequency is pulled towards the driving frequency.

While the mechanism of nonlinearity differs, it is nonetheless clear that the oscillations created by the interaction of the hotspot with an external shunt are capable of similar modulation as observed in Josephson junctions. The generation of mixing products is also similar to the result achieved with hot electron bolometers, which rely on the interference of a weak signal with a local oscillator to create an intermediate frequency signal that drives a thermally induced resistance change in a superconducting nanobridge upon absorption [23],[24]. By getting pulled towards the drive frequency until eventually locking, we observe that relaxation oscillations in the shunted nanowire circuit produce time-averaged characteristics that resemble the AC Josephson effect (see Fig. 4) without requiring a periodic current-phase relationship.

## IV. CONCLUSIONS

In summary, we have presented a study of the nonlinear dynamics in shunted nanowires, and investigated their interaction with periodic external signals. While the non-coherent relaxation oscillations originate quite differently than the famous Josephson oscillations, we find that they are capable of displaying similar time-averaged characteristics and microwave responses as observed in tunnel junctions due to their fast time constants and ability to be synchronized. This has been summarized in our work by two main experimental findings: (1) we have shown that relaxation oscillations may still produce a non-hysteretic current-voltage characteristic that can be fit to the RCSJ model for an overdamped junction, despite the presence of a hotspot; (2) microwave modulation of these oscillations revealed pulling and mixing of the oscillation frequency with the application of a weak external drive, producing steps in the IVC that mimic the AC Josephson effect without matching the expected Shapiro voltage.

Due to the phenomenological similarities observed between the hotspot relaxation oscillation and the Josephson oscillation, we envision that we can use these effects to develop new nanowire-based microwave devices in applications where Josephson oscillations are conventionally employed. While the hotspot relaxation oscillation is incoherent and slower than the Josephson oscillation, it maintains certain advantages such as insensitivity to magnetic noise or vortex dynamics, and it can be fully designed in a circuit model without introducing the complexities of a current-phase relationship. New nanowire devices could be built with these benefits in mind; for example, tuning the oscillation frequency through different shunt resistances may allow for frequency multiplexing, while prior work on tunnel junctions suggests relaxation oscillations may also be used in parametric amplification [10]. Further reducing the inductance may also suppress the hotspot more efficiently, potentially allowing the nanowire to operate using faster, coherent oscillations more similar to those of a Josephson junction. Future studies on the gain and speed of these systems are required to recognize the feasibility of such applications.

**Acknowledgements**

The authors thank Di Zhu, Andrew Dane, Brenden Butters, Marco Colangelo, Cyprian Lewandowski, Prof. Leonid Levitov, and Prof. Daniel Santavicca for scientific discussions, and Dr. Reza Baghdadi for scientific discussion and proof reading of the manuscript. They would also like to thank James Daley and Mark Mondol of the MIT Nanostructures Laboratory for technical support. This research is based on work supported by Intel, and by the Office of the Director of National Intelligence (ODNI), Intelligence Advanced Research Projects Activity (IARPA), via contract W911NF-14-C0089. The views and conclusions contained herein are those of the authors and should not be interpreted as necessarily representing the official policies or endorsements, either expressed or implied, of the ODNI, IARPA, or the U.S. Government. The U.S. Government is authorized to reproduce and distribute reprints for Governmental purposes notwithstanding any copyright annotation thereon. Emily Toomey was supported by the National Science Foundation Graduate Research Fellowship Program (NSF GRFP) under Grant No. 1122374.

**Author Contributions**

E. T. fabricated the nanowires, A.M. supported the superconducting films and fabricated the gold contact pads, E.T. and Q.-Y. Z. took the measurements, E.T. analyzed the data and wrote the paper with input from K.B., Q.-Y.Z., and A.M., and K.B. supervised the project.